\documentclass[
    ,final            
  ]
  {aipproc}

\layoutstyle{6x9}

\def\aj{AJ}%
\def\araa{ARA\&A}%
\def\apj{ApJ}%
\def\apjl{ApJ}%
\def\aap{A\&A}%
\def\mnras{MNRAS}%
\def\nat{Nature}%
\begin{document}

\title{Mass loss and very low-metallicity stars}


\author{Raphael Hirschi}{
  address={Keele University, Lennard-Jones Lab., Keele, ST5 5BG, UK}
}

\author{Cristina Chiappini}{
  address={Osservatorio Astronomico di Trieste, Via G. B. Tiepolo 11, I - 34131 Trieste, Italia}
  ,altaddress={Geneva Observatory, University of Geneva, CH - 1290 Sauverny, Switzerland}
}

\author{Georges Meynet}{
  address={Geneva Observatory, University of Geneva, CH - 1290 Sauverny, Switzerland}
}

\author{Sylvia Ekstr\"om}{
  address={Geneva Observatory, University of Geneva, CH - 1290 Sauverny, Switzerland}
}

\author{Andr\'e Maeder}{
  address={Geneva Observatory, University of Geneva, CH - 1290 Sauverny, Switzerland}
}

\begin{abstract}
Mass loss plays a dominant role in the evolution of massive stars at solar metallicity. After
discussing different mass loss mechanisms and their metallicity dependence, we present the
possibility of strong mass loss at very low metallicity. 
Our models at $Z=10^{-8}$ show that stars more massive than about 60\,$M_\odot$ may lose a
significant fraction of their initial mass in the red supergiant phase. 
This mass loss is due to
the surface enrichment in CNO elements via rotational and convective
mixing. Our 85 $M_\odot$ model ends its life as a fast rotating WO type
Wolf-Rayet star. Therefore the models predict the existence of type Ic SNe and 
long and soft GRBs at very low metallicities. Such strong mass loss in the red supergiant 
phase or the $\Omega \Gamma$-limit
could prevent the most massive stars from ending as pair-creation supernovae.

The very low metallicity models calculated are also very interesting from the nucleosynthesis
point of view. Indeed, the wind of the massive star models can reproduce the CNO 
abundances of the most metal-poor carbon-rich star known to
date, HE1327-2326. Finally, using chemical evolution models, we are able to reproduce the evolution of
CNO elements as observed in the normal extremely metal poor stars.

\end{abstract}

\maketitle


\section{Introduction}
The first generations of stars took part in the re-ionisation of the
universe at the end of the dark ages (roughly 400 million years after the Big Bang).
They are therefore tightly linked to the formation of the first structures in the
universe and can provide valuable information on the early evolution of the
universe. The first stars are thought to be more massive than solar metallicity stars 
\citep{BL04,SOIF06} and mass loss is expected to be very low at very low
metallicities. The logical deduction from these two arguments is that a large fraction of 
the first stars were
very massive at their death ($>$ 100 $M_\odot$) and therefore lead to the production of
pair-creation supernovae (PCSNe). Unfortunately,
the first massive stars died
a long time ago and will probably never be detected directly 
\citep[see however][]{SMWH05,TFS07}.
There are nevertheless indirect observational constraints on the first stars coming
from observations of the most metal poor halo stars \citep{BC05}. 
These observations do not show the peculiar chemical signature of PCSNe 
\citep[strong odd-even effects and low zinc, see][]{HW02}. 
This probably means that at most only a few of these very massive stars (>100 $M_\odot$) 
formed or that they lost a lot of mass even though their initial metal content was very low. In this
paper, we discuss the possibility of strong mass loss at very
low metallicities.

\section{Mass loss and its dependence on metallicity}
At solar metallicity ($Z_\odot$), mass loss has a crucial impact on the evolution of massive stars. It
affects evolutionary tracks, lifetimes and surface abundances. It also
determines the population of massive stars (number of stars in each
Wolf-Rayet subtype for example). It influences the type of supernova at
the death of the star (SNII, Ib, Ic) 
and the final remnant 
(neutron star or black hole). Mass loss releases 
matter, containing newly produced helium and metals
\citep[see][for the impact of mass loss on yields]{CL06},
 and energy back into the interstellar 
medium in amounts comparable
to supernovae (for stars above 30 $M_\odot$). Finally, it affects the hardness of
the ionizing radiation coming from massive stars. It is therefore very
important to understand mass loss in order to understand and model the
evolution of massive stars.

The metallicity ($Z$) dependence of mass loss rates is usually described using the 
formula: 
\begin{equation}
\dot{M}(Z)=\dot{M}(Z_\odot)(Z/Z_\odot)^{\alpha} \label{hirschi:mdot}
\end{equation}
The exponent $\alpha$ varies between 0.5-0.6 
\citep{KP00ARAA,Ku02}
and
0.7-0.86 \citep{VKL01, VdK05} for O-type and WR stars respectively
\citep[See][ for a recent comparison between mass loss
prescriptions and observed mass loss rates]{MKV07}. Until very recently, 
most models use at best the total metal content present at the surface of the star 
to determine the mass loss rate. However, the surface chemical composition becomes
very different from the solar mixture, due either to mass loss in the WR stage or by
internal mixing (convection and rotation) after the main sequence. It is
therefore important to know the contribution from each chemical species to opacity and
mass loss. Recent studies \citep{VKL00,VdK05} show that iron is the dominant
element concerning radiation line-driven mass loss for O-type and WR stars. 
In the case of WR stars, there is
however a plateau at low metallicity due to the contributions from light elements
like carbon, nitrogen and oxygen (CNO). 
In the red supergiant (RSG) stage, the rates generally used are still those 
of \citet{NdJ90}. More recent observations indicate that
there is a very weak dependence of dust-driven mass loss on metallicity and 
that CNO elements and especially nucleation seed components like silicon and titanium are
dominant \citep{VL00,VL06,FG06}. 
See \citet{VL05} for recent mass loss rate prescriptions in the 
RSG stage. 
In particular, the ratio of carbon to oxygen is important to determine which
kind of molecules and dusts form. 
If the ratio of carbon to oxygen is larger than one, then carbon-rich dust would
form, and more likely drive a wind since they are more opaque than oxygen-rich
dust at low metallicity \citep{HA07}. 

In between the hot and cool parts of the HR-diagram, mass
loss is not well understood. Observations of the LBV stage indicate that several solar
masses per year may be lost \citep{Sm03} and there is no indication of a metallicity dependence.
Chromospheric activity could also play a role in stars having surface temperature
similar to the Sun. 
Thermally driven winds \citep{SC07} and pulsations are still other ways to lose mass.

The mass loss prescriptions used in the Geneva stellar evolution code are described in detail in
\citet{ROTXI}. In particular, the mass loss rates depend 
on metallicity as $\dot{M} \sim (Z/Z_{\odot})^{0.5}$, where
$Z$ is the mass fraction of heavy elements at the surface
of the star. Models including the effects of both mass loss and especially rotation better 
reproduce the WR/O
ratio and also the ratio of type Ib+Ic to type II supernova as a function of metallicity compared
to non-rotating models, which underestimate these ratios \citep{ROTXI}. 
Rotating models also better reproduce the ratio of blue to red supergiants at low metallicities 
and surface enrichments in nitrogen during the main sequence \citep{MM00}. 

\section{First stellar generations}
As we saw in the previous sections, mass loss plays a crucial role in the evolution of
solar metallicity stars. In this section, we discuss the importance of mass
loss on the evolution of the first stellar generations. 
The first stellar
generations are different from solar metallicity stars due to their low metal content
or absence of it. First, very low-$Z$ stars are more compact due to lower opacity.
Second, {\it metal free} stars burn hydrogen in a core, which is denser
and hotter due to the lack of initial CNO elements. 
This implies that the transition between core hydrogen and helium burning is much
shorter and smoother. Furthermore, hydrogen burns via
the pp-chain in shell burning. {\it These differences make the metal free (first) stars different from the
second or later generation stars!} \citep{EM07}. Third, mass loss is
metallicity dependent (at least for radiation-driven winds) and therefore is expected to 
become very weak at very
low metallicity. Finally, the initial mass function of the first stellar generations is
expected to be top heavy below a certain threshold 
\citep{BL03}. 

Mass loss is expected to be very weak. What could change this expectation? 
In order to investigate this question, models at very low and zero metallicity were calculated
with the same physics as in the models able to reproduce many observables around solar
metallicity. 
These models were computed with an 
initial total angular momentum similar or slightly higher to the one contained in solar 
metallicity models with average initial rotational velocities of 300\,km\,s$^{-1}$.
Models of metal free stars
including the effect of rotation \citep{EMM05} show that stars may lose up to
10 \% of their initial mass due to the star rotating at its critical limit (also
called break-up limit). 
The mass loss due to the star reaching the critical
limit is non-negligible but not important enough to change
drastically the fate of the first generation stars. 

\begin{figure}[ht]
\includegraphics[height=.3\textheight]{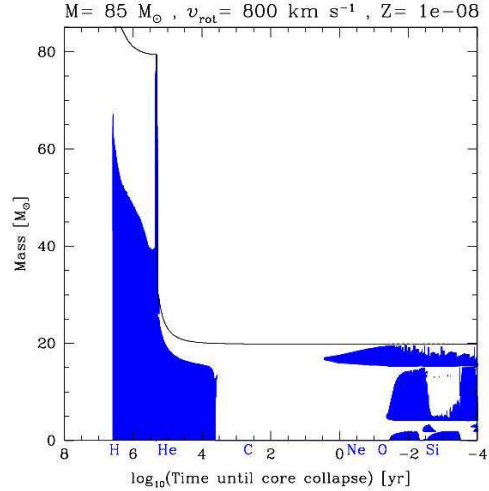}
\caption{Structure evolution diagram of the 85 $M_\odot$ model at $Z=10^{-8}$.
Coloured areas correspond to convective zones along the lagrangian mass coordinate as a
function of the time left until the core collapse. The top solid line shows the total mass of the
star. The burning stage abbreviations are given below the time axis.
}
\label{hirschi:kip85}
\end{figure}

The situation is very different at very low but non-zero metallicity 
\citep{MEM06,H07}. 
The total mass of an 85\,$M_\odot$ model at $Z=10^{-8}$ is shown in Fig.
\ref{hirschi:kip85} by the top solid line. This model, like metal free models,
loses around 5\% of its initial mass when its surface reaches break-up velocities in
the second part of the main sequence. At the end of core H-burning,
the core contracts and the envelope expands, thus decreasing the surface
velocity and its ratio to the critical velocity. The mass loss rate becomes
very low again until the star crosses the HR diagram and reaches the RSG
stage. At this point the convective envelope dredges up CNO elements to
the surface increasing its overall metallicity. The total metallicity, $Z$, is 
used in this model (including CNO elements)
for the metallicity dependence of the mass loss.
Therefore depending on how much CNO is brought up to the surface, the
mass loss becomes very large again. The CNO brought to the surface
comes from primary C and O produced in He-burning. Rotational and convective 
mixing brings these elements
into the H-burning shell. A large fraction of the C and O is then 
transformed into primary nitrogen via the CNO cycle. 
Additional convective and rotational
mixing is necessary to bring the primary CNO to the surface of the star.

The strongest mass loss occurs in these models in the cooler part of the HR
diagram. Dust-driven winds appear to be metallicity independent as long as C-rich dust
can form. For this to occur, the surface effective temperature needs to be low enough
(log(T$_{\rm eff})<3.6$) and carbon needs to be more abundant than oxygen. Note that
nucleation seeds (probably involving titanium) are still necessary to form C-rich dust. 
It is not clear
whether extremely low-$Z$ stars will reach such low effective temperatures. This
depends on the opacity and the opacity tables used in our calculations did not account for the
non-standard mixture of metals 
\citep[high CNO and low iron abundance, see][ for possible effects]{Ma02}. 
It is interesting to note that the wind of the 85 $M_\odot$ model
is richer in carbon than oxygen, thus allowing C-rich dust to form if nucleation seeds are
present. There may also be
other important types of wind, like Chromospheric activity-driven, pulsation-driven, 
thermally-driven or continuum-driven winds.

\section{Nucleosynthesis and chemical evolution}
\subsection{The most metal poor star known to date, HE1327-2326}

\begin{figure}[ht]
\includegraphics[height=.3\textheight]{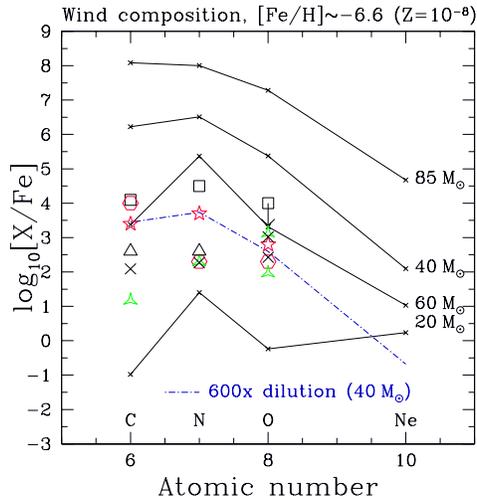}
\caption{Composition in [X/Fe] of the stellar wind for the $Z=10^{-8}$
models (solid lines).
For HE1327-2326 ({\it red stars}), the best fit for the
CNO elements is
obtained by diluting the composition of the wind of the 40 $M_\odot$
model by a factor 600 \citep[see][ for more details]{H07}.}
\label{hirschi:cempw}
\end{figure}
Significant mass loss in very low-$Z$ massive stars offers an interesting
explanation for the strong enrichment in CNO elements of the most metal
poor stars observed in the halo of the galaxy 
\citep[see][]{MEM06,H07}. 
The most metal poor stars known to date, 
HE1327-2326 \citep{Fr06} is characterised by very high N, C and O abundances,
high Na, Mg and Al abundances, a weak s--process enrichment and depleted
lithium. The star is not evolved so has not had time to bring
self--produced CNO elements to its surface and is most likely a subgiant.
By using one or a few SNe and using a very large mass cut, 
\citet{LCB03} and \citet{IUTNM05} are
able to reproduce the abundance of most elements. 
However they are not
able to reproduce the nitrogen surface abundance of
HE1327-2326 without rotational mixing. 
A lot of the features of this star are similar to the properties of the
stellar winds of very metal poor rotating stars. 
HE1327-2326 may therefore have formed
from gas, which was mainly enriched by stellar winds of rotating very low
metallicity stars. In this scenario, a first generation of stars 
(PopIII) 
pollutes the interstellar medium to very low metallicities
([Fe/H]$\sim$-6). Then a PopII.5 star 
\citep{paris05} like the 
40 $M_\odot$ model calculated here
pollutes (mainly through its wind) the interstellar medium out of
 which HE1327-2326 forms.
This would mean that HE1327-2326 is a third generation star.
In this scenario, 
the CNO abundances are well reproduced, in particular that of
nitrogen, which according to the new values for a subgiant from \citet{Fr06}
is 0.9 dex higher in [X/Fe] than oxygen. 
This is shown in Fig. \ref{hirschi:cempw} where the abundances of HE1327-2326 are
represented by the red stars and the best fit is 
obtained by diluting the composition of the wind of the 40 $M_\odot$
model by a factor 600. When the SN
contribution is added, the [X/Fe] ratio is usually lower for nitrogen
than for oxygen. 
%
It is interesting to note that the very high CNO yields of the 
40 $M_\odot$ stars brings the total
metallicity $Z$ above the limit for low mass star formation
obtained in \citet{BL03}.

\subsection{Primary nitrogen}

\begin{figure}
  \includegraphics[height=.3\textheight]{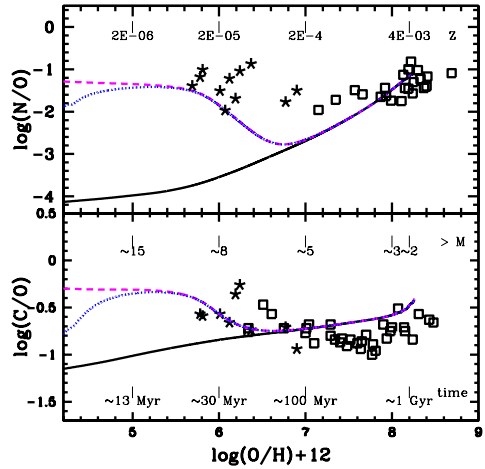}
  \caption{Chemical evolution model predictions of the N/O and C/O
evolution, in the galactic
halo, for different stellar evolution inputs. 
The solid curves show the predictions of a model without
fast rotators at low metallicities. The dashed and dotted lines,
almost overlapping, show the effect of including a population of fast
rotators at low metallicities (the dotted line includes also the Z=0 
fast rotators). 
For the data see \cite{CH06} and references therein.}
\label{CNO}
\end{figure}

One of the most stringent observational constraint at very low Z is a very
high primary $^{14}$N production.
This requires extremely high primary $^{14}$N production in massive
stars, about 0.1 $M_\odot$ per star \citep[$\sim$0.15 $M_\odot$ used 
in the heuristic model
of][]{CMB05}.
Upon the inclusion of the new stellar calculations of
\citet{H07} for Z$=$10$^{-8}$
in a chemical evolution model for the galactic halo with infall and outflow, 
both high N/O and C/O ratios are obtained in the very-metal poor
metallicity range in agreement with observations 
\citep[see details in][]{CH06b}. This model is shown in Fig. \ref{CNO}
(dashed magenta curve). In the same figure, a model computed without
fast rotators (solid black curve) is also shown. Fast rotation
enhances the nitrogen production by $\sim$3 orders of magnitude.
These results also offer a natural explanation for the large 
scatter observed in the N/O abundance ratio of {\it normal} metal-poor 
halo stars: given the strong dependency of the nitrogen yields on the 
rotational velocity of the star, we expect a
scatter in the N/O ratio which could be the consequence of the distribution 
of the stellar rotational velocities as a function of metallicity. 
Furthermore, the strong production of primary nitrogen
is linked to a very active H-burning shell which contributes a large part
of the total luminosity of the star. Hence, the energy produced by the
helium core is reduced, reducing the efficiency of the 
$^{12}$C($\alpha$,$\gamma$)$^{16}$O reaction. 
As a consequence, less carbon is turned 
into oxygen, producing high C/O ratios.
Although the abundance data for C/O is still very uncertain, 
a C/O upturn at low
metallicities is suggested by observations \citep[see][ and references therein]{A05}.

\section{Gamma-ray bursts and pair-creation supernovae}
Long and soft gamma-ray bursts (GRBs) have now been firmly connected to the death
of type Ic supernovae \citep[see][ for a recent review]{WB06}. In one of
the most promising models, the collapsar model \citep{W93}, GRB progenitors
must form a black hole, lose their hydrogen rich envelope (become a WR) and retain enough 
angular momentum in their core during the pre-supernova stages. 
The strong mass loss discussed in the previous section makes it possible
for single massive stars in the first stellar generations to become WR stars and
even to retain enough angular momentum to produce a GRB \citep{H07}. 
A wide grid of models around $Z_\odot$ shows that the
fast rotating WO stars are produced only at metallicities equal to or lower than that of 
the LMC, in agreement with observations \citep{St06}.
These models predict that, at low $Z$, one GRB occurs for every 10 core collapse 
supernovae, which is in the upper limit of what is allowed by the observations  
\citep[see][ for more details]{grb05}.
More recent models
including the effects of magnetic fields \citep{YL05} show that another mechanism is
possible to produce GRBs at low $Z$. This mechanism is the quasi-homogeneous
evolution of very fast rotating massive stars. In this scenario, WR stars are
produced by mixing and not mass loss. This last scenario however does not predict
GRB at metallicities equal or higher than the SMC. This upper limit is too low
compared to recent observations \citep{St06}. 
Taking into account the anisotropy in the wind of these fast rotating stars 
\citep{AM07}
may help reduce the discrepancy between models and observations. 
Note that the 
downward revision of solar metallicity \citep{A05} 
may help resolve the problem.

Apart from GRBs, pair-creation supernovae (PCSNe) are very energetic explosions,
which could be observed up to very high redshifts 
\citep{SMWH05}.
PCSNe are expected
to follow the death of stars in the mass range between 100 and 250 $M_\odot$, assuming
that they do not lose a significant fraction of their mass during the
pre-supernova stages. Amongst the very first stars formed in the Universe, one
expects to have PCSN due to the lack of mass loss and to the low opacity
unable to stop the accretion on the star during its formation.
However, the EMP stars observed in the halo of the galaxy do not show the
peculiar chemical signature of PCSN \citep[strong odd-even effect, see][]{HW02}. 
This means that either too few or even no 
PCSN existed. One possible explanation to avoid the production of very
low-$Z$ or metal free PCSNe is the strong mass loss in the cool part of the HR
diagram due to the surface enrichment in CNO elements induced by rotational and 
convective mixing (see previous section) or the star reaching the 
$\Omega \Gamma$-limit \citep{EM07}.

\section{Conclusion}
For the most massive models ($M > 60$\,$M_\odot$),
significant mass
loss occurs during the red supergiant stage assuming that CNO elements
are important contributors to mass loss. This mass loss is due to
the surface enrichment in CNO elements via rotational and convective
mixing. 
The models predict the production of WR stars for an initial mass higher
than 60 $M_\odot$ at $Z=10^{-8}$ and our 85 $M_\odot$ model with 
$\upsilon_{\rm ini}=800$\,km\,s$^{-1}$ becomes a WO.
Therefore SNe of type Ib and Ic are predicted from single massive stars
at these low metallicities. The 85
$M_\odot$ model retains enough angular momentum to produce a GRB but
the calculations did not include the effects of magnetic fields.

The stellar yields were calculated for light elements. These yields 
were used in a galactic 
chemical evolution model and successfully reproduce the early 
evolution of CNO elements \citep{CH06}. Finally, a scenario is proposed to
explain the CNO abundances of the most metal poor star known to date 
HE1327-2326. 



\bibliographystyle{mn2e}   

\begin{thebibliography}{}

\bibitem[\protect\citeauthoryear{{Asplund}}{{Asplund}}{2005}]{A05}
{Asplund} M.,  2005, \araa, 43, 481

\bibitem[\protect\citeauthoryear{{Beers} \& {Christlieb}}{{Beers} \&
  {Christlieb}}{2005}]{BC05}
{Beers} T.~C.,  {Christlieb} N.,  2005, \araa, 43, 531

\bibitem[\protect\citeauthoryear{{Bromm} \& {Larson}}{{Bromm} \&
  {Larson}}{2004}]{BL04}
{Bromm} V.,  {Larson} R.~B.,  2004, \araa, 42, 79

\bibitem[\protect\citeauthoryear{{Bromm} \& {Loeb}}{{Bromm} \&
  {Loeb}}{2003}]{BL03}
{Bromm} V.,  {Loeb} A.,  2003, \nat, 425, 812

\bibitem[\protect\citeauthoryear{{Chiappini}, {Hirschi}, {Matteucci}, {Meynet},
  {Ekstroem} \& {Maeder}}{{Chiappini} et~al.}{2006b}]{CH06b}
{Chiappini} C.,  {Hirschi} R.,  {Matteucci} F.,  {Meynet} G.,  {Ekstroem} S.,
   {Maeder} A.,  2006b, in Proceedings of Nuclei in the Cosmos IX, CERN,
  PoS(NIC-IX)080

\bibitem[\protect\citeauthoryear{{Chiappini}, {Hirschi}, {Meynet},
  {Ekstr{\"o}m}, {Maeder} \& {Matteucci}}{{Chiappini} et~al.}{2006}]{CH06}
{Chiappini} C.,  {Hirschi} R.,  {Meynet} G.,  {Ekstr{\"o}m} S.,  {Maeder} A.,
   {Matteucci} F.,  2006, \aap, 449, L27

\bibitem[\protect\citeauthoryear{{Chiappini}, {Matteucci} \&
  {Ballero}}{{Chiappini} et~al.}{2005}]{CMB05}
{Chiappini} C.,  {Matteucci} F.,    {Ballero} S.~K.,  2005, \aap, 437, 429

\bibitem[\protect\citeauthoryear{{Chieffi} \& {Limongi}}{{Chieffi} \&
  {Limongi}}{2006}]{CL06}
{Chieffi} A.,  {Limongi} M.,  2006, in Proceedings of Nuclei in the Cosmos IX,
  CERN, PoS(NIC-IX)250

\bibitem[\protect\citeauthoryear{{Ekstr{\"o}m}, {Meynet} \&
  {Maeder}}{{Ekstr{\"o}m} et~al.}{2006}]{EMM05}
{Ekstr{\"o}m} S.,  {Meynet} G.,    {Maeder} A.,  2006, in {Lamers} H.,
  {Langer} N.,  {Nugis} T.,   {Annuk} K.,  eds, Stellar Evolution at Low
  Metallicity: Mass Loss, Explosions, Cosmology Vol.~353 of ASP Conf. Series.
p.~141

\bibitem[\protect\citeauthoryear{{Ekstr{\"o}m}, {Meynet} \&
  {Maeder}}{{Ekstr{\"o}m} et~al.}{2007}]{EM07}
{Ekstr{\"o}m} S.,  {Meynet} G.,    {Maeder} A.,  2007, ArXiv e-prints0709.0202,
  709

\bibitem[\protect\citeauthoryear{{Ferrarotti} \& {Gail}}{{Ferrarotti} \&
  {Gail}}{2006}]{FG06}
{Ferrarotti} A.~S.,  {Gail} H.-P.,  2006, \aap, 447, 553

\bibitem[\protect\citeauthoryear{{Frebel}, {Christlieb}, {Norris}, {Aoki} \&
  {Asplund}}{{Frebel} et~al.}{2006}]{Fr06}
{Frebel} A.,  {Christlieb} N.,  {Norris} J.~E.,  {Aoki} W.,    {Asplund} M.,
  2006, \apjl, 638, L17

\bibitem[\protect\citeauthoryear{{Heger} \& {Woosley}}{{Heger} \&
  {Woosley}}{2002}]{HW02}
{Heger} A.,  {Woosley} S.~E.,  2002, \apj, 567, 532

\bibitem[\protect\citeauthoryear{{Hirschi}}{{Hirschi}}{2005}]{paris05}
{Hirschi} R.,  2005, in {Hill} V.,  {Fran{\c c}ois} P.,   {Primas} F.,  eds,
  IAU Symposium 228 {PopII 1/2 stars: very high 14N and low 16O yields}.
pp 331--332

\bibitem[\protect\citeauthoryear{{Hirschi}}{{Hirschi}}{2007}]{H07}
{Hirschi} R.,  2007, \aap, 461, 571

\bibitem[\protect\citeauthoryear{{Hirschi}, {Meynet} \& {Maeder}}{{Hirschi}
  et~al.}{2005}]{grb05}
{Hirschi} R.,  {Meynet} G.,    {Maeder} A.,  2005, \aap, 443, 581

\bibitem[\protect\citeauthoryear{{H{\"o}fner} \& {Andersen}}{{H{\"o}fner} \&
  {Andersen}}{2007}]{HA07}
{H{\"o}fner} S.,  {Andersen} A.~C.,  2007, \aap, 465, L39

\bibitem[\protect\citeauthoryear{{Iwamoto}, {Umeda}, {Tominaga}, {Nomoto} \&
  {Maeda}}{{Iwamoto} et~al.}{2005}]{IUTNM05}
{Iwamoto} N.,  {Umeda} H.,  {Tominaga} N.,  {Nomoto} K.,    {Maeda} K.,  2005,
  Science, 309, 451

\bibitem[\protect\citeauthoryear{{Kudritzki}}{{Kudritzki}}{2002}]{Ku02}
{Kudritzki} R.~P.,  2002, \apj, 577, 389

\bibitem[\protect\citeauthoryear{{Kudritzki} \& {Puls}}{{Kudritzki} \&
  {Puls}}{2000}]{KP00ARAA}
{Kudritzki} R.-P.,  {Puls} J.,  2000, \araa, 38, 613

\bibitem[\protect\citeauthoryear{{Limongi}, {Chieffi} \& {Bonifacio}}{{Limongi}
  et~al.}{2003}]{LCB03}
{Limongi} M.,  {Chieffi} A.,    {Bonifacio} P.,  2003, \apjl, 594, L123

\bibitem[\protect\citeauthoryear{{Maeder} \& {Meynet}}{{Maeder} \&
  {Meynet}}{2000}]{MM00}
{Maeder} A.,  {Meynet} G.,  2000, \araa, 38, 143

\bibitem[\protect\citeauthoryear{{Marigo}}{{Marigo}}{2002}]{Ma02}
{Marigo} P.,  2002, \aap, 387, 507

\bibitem[\protect\citeauthoryear{{Meynet}, {Ekstr{\"o}m} \& {Maeder}}{{Meynet}
  et~al.}{2006}]{MEM06}
{Meynet} G.,  {Ekstr{\"o}m} S.,    {Maeder} A.,  2006, \aap, 447, 623

\bibitem[\protect\citeauthoryear{{Meynet} \& {Maeder}}{{Meynet} \&
  {Maeder}}{2005}]{ROTXI}
{Meynet} G.,  {Maeder} A.,  2005, \aap, 429, 581

\bibitem[\protect\citeauthoryear{{Meynet} \& {Maeder}}{{Meynet} \&
  {Maeder}}{2007}]{AM07}
{Meynet} G.,  {Maeder} A.,  2007, \aap, 464, L11

\bibitem[\protect\citeauthoryear{{Mokiem}, {de Koter}, {Vink}, {Puls}, {Evans},
  {Smartt}, {Crowther}, {Herrero}, {Langer}, {Lennon}, {Najarro} \&
  {Villamariz}}{{Mokiem} et~al.}{2007}]{MKV07}
{Mokiem} M.~R.,  {de Koter} A.,  {Vink} J.~S.,  {Puls} J.,  {Evans} C.~J.,
  {Smartt} S.~J.,  {Crowther} P.~A.,  {Herrero} A.,  {Langer} N.,  {Lennon}
  D.~J.,  {Najarro} F.,    {Villamariz} M.~R.,  2007, ArXiv e-prints0708.2042,
  708

\bibitem[\protect\citeauthoryear{{Nieuwenhuijzen} \& {de
  Jager}}{{Nieuwenhuijzen} \& {de Jager}}{1990}]{NdJ90}
{Nieuwenhuijzen} H.,  {de Jager} C.,  1990, \aap, 231, 134

\bibitem[\protect\citeauthoryear{{Scannapieco}, {Madau}, {Woosley}, {Heger} \&
  {Ferrara}}{{Scannapieco} et~al.}{2005}]{SMWH05}
{Scannapieco} E.,  {Madau} P.,  {Woosley} S.,  {Heger} A.,    {Ferrara} A.,
  2005, \apj, 633, 1031

\bibitem[\protect\citeauthoryear{{Schneider}, {Omukai}, {Inoue} \&
  {Ferrara}}{{Schneider} et~al.}{2006}]{SOIF06}
{Schneider} R.,  {Omukai} K.,  {Inoue} A.~K.,    {Ferrara} A.,  2006, \mnras,
  369, 1437

\bibitem[\protect\citeauthoryear{{Schr{\"o}der} \& {Cuntz}}{{Schr{\"o}der} \&
  {Cuntz}}{2007}]{SC07}
{Schr{\"o}der} K.-P.,  {Cuntz} M.,  2007, \aap, 465, 593

\bibitem[\protect\citeauthoryear{{Smith}, {Gehrz}, {Hinz}, {Hoffmann}, {Hora},
  {Mamajek} \& {Meyer}}{{Smith} et~al.}{2003}]{Sm03}
{Smith} N.,  {Gehrz} R.~D.,  {Hinz} P.~M.,  {Hoffmann} W.~F.,  {Hora} J.~L.,
  {Mamajek} E.~E.,    {Meyer} M.~R.,  2003, \aj, 125, 1458

\bibitem[\protect\citeauthoryear{{Stanek}, {Gnedin}, {Beacom}, {Gould},
  {Johnson}, {Kollmeier}, {Modjaz}, {Pinsonneault}, {Pogge} \&
  {Weinberg}}{{Stanek} et~al.}{2006}]{St06}
{Stanek} K.~Z.,  {Gnedin} O.~Y.,  {Beacom} J.~F.,  {Gould} A.~P.,  {Johnson}
  J.~A.,  {Kollmeier} J.~A.,  {Modjaz} M.,  {Pinsonneault} M.~H.,  {Pogge} R.,
    {Weinberg} D.~H.,  2006, Acta Astronomica, 56, 333

\bibitem[\protect\citeauthoryear{{Tornatore}, {Ferrara} \&
  {Schneider}}{{Tornatore} et~al.}{2007}]{TFS07}
{Tornatore} L.,  {Ferrara} A.,    {Schneider} R.,  2007, ArXiv
  e-prints0707.1433, 707

\bibitem[\protect\citeauthoryear{{van Loon}}{{van Loon}}{2000}]{VL00}
{van Loon} J.~T.,  2000, \aap, 354, 125

\bibitem[\protect\citeauthoryear{{van Loon}}{{van Loon}}{2006}]{VL06}
{van Loon} J.~T.,  2006, in {Lamers} H.~J.~G.~L.~M.,  {Langer} N.,  {Nugis} T.,
    {Annuk} K.,  eds, ASP Conf. Ser. 353: Stellar Evolution at Low Metallicity:
  Mass Loss, Explosions, Cosmology p.~211

\bibitem[\protect\citeauthoryear{{van Loon}, {Cioni}, {Zijlstra} \&
  {Loup}}{{van Loon} et~al.}{2005}]{VL05}
{van Loon} J.~T.,  {Cioni} M.-R.~L.,  {Zijlstra} A.~A.,    {Loup} C.,  2005,
  \aap, 438, 273

\bibitem[\protect\citeauthoryear{{Vink} \& {de Koter}}{{Vink} \& {de
  Koter}}{2005}]{VdK05}
{Vink} J.~S.,  {de Koter} A.,  2005, \aap, 442, 587

\bibitem[\protect\citeauthoryear{{Vink}, {de Koter} \& {Lamers}}{{Vink}
  et~al.}{2000}]{VKL00}
{Vink} J.~S.,  {de Koter} A.,    {Lamers} H.~J.~G.~L.~M.,  2000, \aap, 362, 295

\bibitem[\protect\citeauthoryear{{Vink}, {de Koter} \& {Lamers}}{{Vink}
  et~al.}{2001}]{VKL01}
{Vink} J.~S.,  {de Koter} A.,    {Lamers} H.~J.~G.~L.~M.,  2001, \aap, 369, 574

\bibitem[\protect\citeauthoryear{{Woosley}}{{Woosley}}{1993}]{W93}
{Woosley} S.~E.,  1993, \apj, 405, 273

\bibitem[\protect\citeauthoryear{{Woosley} \& {Bloom}}{{Woosley} \&
  {Bloom}}{2006}]{WB06}
{Woosley} S.~E.,  {Bloom} J.~S.,  2006, \araa, 44, 507

\bibitem[\protect\citeauthoryear{{Yoon} \& {Langer}}{{Yoon} \&
  {Langer}}{2005}]{YL05}
{Yoon} S.-C.,  {Langer} N.,  2005, \aap, 443, 643

\end{thebibliography}

\end{document}